\documentclass[conference]{IEEEtran}
\IEEEoverridecommandlockouts
\usepackage[T1]{fontenc}
\usepackage[utf8]{inputenc}
\usepackage{cite}
\usepackage{amsmath,amssymb,amsfonts}
\usepackage{algorithmic}
\usepackage{algorithm}
\usepackage{graphicx}
\usepackage{textcomp}
\usepackage{tabularx}
\usepackage{xcolor}
\usepackage{bbm}
\usepackage{graphicx}
\usepackage{adjustbox}
\usepackage{hyperref}
\usepackage{type1cm} % If using scalable fonts
\def\BibTeX{{\rm B\kern-.05em{\sc i\kern-.025em b}\kern-.08em
    T\kern-.1667em\lower.7ex\hbox{E}\kern-.125emX}}
\begin{document}

\title{Physical Backdoor Attack Against Deep Learning-based Modulation Classification%*\\
%{\footnotesize \textsuperscript{*}Note: Sub-titles are not captured for https://ieeexplore.ieee.org  and
%should not be used}
\thanks{The presented work has been funded in whole by the National Science Centre, Poland, within grant no. 2023/05/Y/ST7/00002 on “Physics-based wireless AI providing scalability and efficiency” (PASSIONATE) within the CHIST-ERA programme. For the purpose of Open Access, the author has applied a CC-BY public copyright license to any Author Accepted Manuscript (AAM) version arising from this submission.}}

\author{\IEEEauthorblockN{Younes Salmi}
\IEEEauthorblockA{\textit{Institute of Radiocommunications} \\
\textit{Poznan University of Technology}\\
Poznań, Poland \\
younes.salmi@put.poznan.pl}
\and
\IEEEauthorblockN{Hanna Bogucka}
\IEEEauthorblockA{\textit{Institute of Radiocommunications} \\
\textit{Poznan University of Technology}\\
Poznań, Poland \\
hanna.bogucka@put.poznan.pl}
}

\maketitle

\begin{abstract}
Deep Learning (DL) has become a key technology that assists radio frequency (RF) signal classification applications, such as modulation classification.
However, the DL models are vulnerable to adversarial machine learning threats, such as data manipulation attacks.
We study a physical backdoor (Trojan) attack that targets a DL-based modulation classifier.
In contrast to digital backdoor attacks, where digital triggers are injected into the training dataset, we use power amplifier (PA) non-linear distortions to create physical triggers before the dataset is formed.
During training, the adversary manipulates amplitudes of RF signals and changes their labels to a target modulation scheme, training a backdoored model.
At inference, the adversary aims to keep the backdoor attack inactive such that the backdoored model maintains high accuracy on test signals. 
However, if they apply the same manipulation used during training on these test signals, the backdoor is activated, and the model misclassifies these signals.
We demonstrate that our proposed attack achieves high attack success rates with few manipulated RD signals for different noise levels.
Furthermore, we test the resilience of the proposed attack to multiple defense techniques, and the results show that these techniques fail to mitigate the attack. 
\end{abstract}


\section{Introduction}
\label{sec:intro}

Deep learning (DL) models have been approved as powerful tools for modeling and managing processes in wireless communications, including signal processing in the physical layer (PHY) \cite{o2017introduction}.
DL-based methods have been applied to the classification of baseband and radio frequency (RF) signals, such as modulation classification and RF fingerprinting.
Artificial Neural Networks (ANNs) extract the complex features of raw RF signals in a high-dimensional space, allowing the classification of modulation schemes from various statistical features \cite{lee2017deep}.
O'Shea et al in \cite{o2016convolutional} show that Convolutional Neural Networks (CNNs) are a strong candidate approach for modulation classification, especially in low Signal-to-Noise Ratio (SNR) regions.
In \cite{mer2019Fing}, a trained recurrent neural network (RNN) authenticates a single transmitting device against a population of other identical devices for RF fingerprinting.

Despite the prominence of DL techniques, they are vulnerable to threats from the class of Adversarial Machine Learning (AML) attacks  \cite{adesina2022adversarial}.
For instance, exploratory attacks aim to discover information on a DL model used, to launch further attacks aimed at fooling the DL algorithm.
In \cite{zhao2024explanation}, the authors trained a CNN surrogate model to discover the candidate data samples to be attacked.
In addition, methods such as the fast gradient sign method (FGSM) \cite{flowers2019evaluating} and Carlini-Wagner (C\&W) \cite{kokalj2019targeted} attacks are used to craft adversarial (evasion) attacks on modulation recognition.
Moreover, causative attacks manipulate the training process by injecting poisons (perturbations) into the dataset; this may lead to classification inaccuracies. 

A new kind of AML attack called backdoor or Trojan attacks \cite{chen2017targeted} is the focus of research addressing the vulnerabilities of DL in the classification of RF signals. 
In this case, the adversary embeds triggers (poisons) in the training dataset (which is a causative attack) and activates them for the targeted classification class at inference (for a successful evasion attack).
Even in the presence of a causative attack,
the attacked DL model maintains classification accuracy if the test data are clean (without a trigger).
However, when the backdoor is activated (test data with triggers), the model becomes biased toward the targeted class.
For instance, adversaries may launch this attack to disrupt the functionality of DL-based receivers that use modulation classification.
When receivers achieve high classification accuracy on clean data, they will be disrupted by triggers and incorrectly classify legitimate signals, leading to interrupted communications.

This paper addresses the backdoor attacks on deep learning models used for modulation classification.
In particular, the design of a physical backdoor attack using the distortions caused by the hardware is considered.
The study focuses on the vulnerability of such DL tools to physical backdoor attacks, while mitigation is kept for future work.

\section{Related Works}
\label{sec:related}

Several Backdoor attacks were proposed in \cite{sagduyu2019Trojan, li2023AMC, Zhang2024Wavelet, vavilapalli2024Input} for modulation classification,
Sagduyu et al. \cite{sagduyu2019Trojan} proposed the first backdoor attacks targeting RF communications.
The authors generate triggers by modifying the phases of IQ samples and changing their labels to a targeted class.
In \cite{li2023AMC}, low-power Gaussian noise was used to modify the amplitude of the IQ samples at random data locations.
A clean-label attack based on Wavelet Domain Frequency Steganography was proposed in \cite{Zhang2024Wavelet} for backdoor stealthiness.
In \cite{vavilapalli2024Input}, the authors used Gram-Schmidt orthogonalization to create two linearly independent random vectors that generate I and Q triggers.
While for RF fingerprinting, authors in \cite{Zhao2024Explanable} discussed a model-agnostic approach. 
It employs explainable artificial intelligence (XAI) and an autoencoder to determine the trigger place and value.

Multiple works \cite{zhao2023Stealthy, huang2023Modulation, huang2023hidden, Xu2024Adaptive} considered designing backdoor attacks for modulation classification and RF fingerprinting simultaneously. 
A transferable backdoor attack between modulation classification and RF fingerprinting was studied in \cite{zhao2023Stealthy}. 
The attack has shown high effectiveness across different signals using a gradient-based optimization.
Clean-label backdoor attacks were proposed in \cite{huang2023Modulation, huang2023hidden} using projected gradient descent (PGD) to balance the effectiveness and stealthiness of the triggers.
Another Xu et al. \cite{Xu2024Adaptive} proposed an adaptive trigger generation that tailors different backdoors for different RF applications using an adversarial generative network (GAN).

To defend DL-based signal classifiers against backdoor attacks, data augmentation was proposed in \cite{adesina2022adversarial}. 
The approach augments the training data with random rotations to introduce backdoor robustness to the classifier.
Another data-level defense is activation clustering \cite{adesina2022adversarial, vavilapalli2024Input, huang2023Modulation} used for modulation classification.
It starts with training models on poisoned datasets, and then the activations of the last hidden layers with dimensionality reduction techniques are applied to separate the clean and poisoned samples.
Neural cleanse defenses that involve reverse engineering of the DL-based RF model were used in \cite{zhao2023Stealthy, Zhang2024Wavelet} to find label manipulations.
\cite{Zhao2024Explanable, Zhang2024Wavelet} discussed fine-pruning defenses. 
These techniques eliminate backdoors by identifying and pruning inactive neurons activated by adversarial backdoor triggers.
STRong Intentional Perturbation (STRIP) was used in \cite{zhao2023Stealthy, Zhang2024Wavelet} to detect backdoored inputs.
This technique uses Shannon entropy, where low entropy (consistent predictions) yields a backdoored input, while high entropy (diverse predictions) yields a clean input to exploit this weakness by intentionally perturbing an input and then observing the consistency of the model predictions.

However, the mentioned works have focused on backdoor attacks that manipulate datasets formed of received RF signals with their modulation schemes.
This is referred to as the \emph{digital} manipulation, where the dataset is manipulated after its formation. 
Such manipulation is often impractical in cyber-secure systems. 
First, accessing the dataset challenges the threat actors to bypass the cybersecurity monitoring systems.
Additionally, the \emph{Recover} function of cybersecurity frameworks may recover a clean version of models and datasets if a suspicious activity, such as accessing the dataset, is detected.
Moreover, the dataset accessed is not always represented in raw samples. 
Normalization methods might be employed to change the representation of samples, making the attacker's task more challenging.

\begin{figure*}[htbp]
    \centering
    \adjustbox{max width=\linewidth}{
    \includegraphics[width=\linewidth]{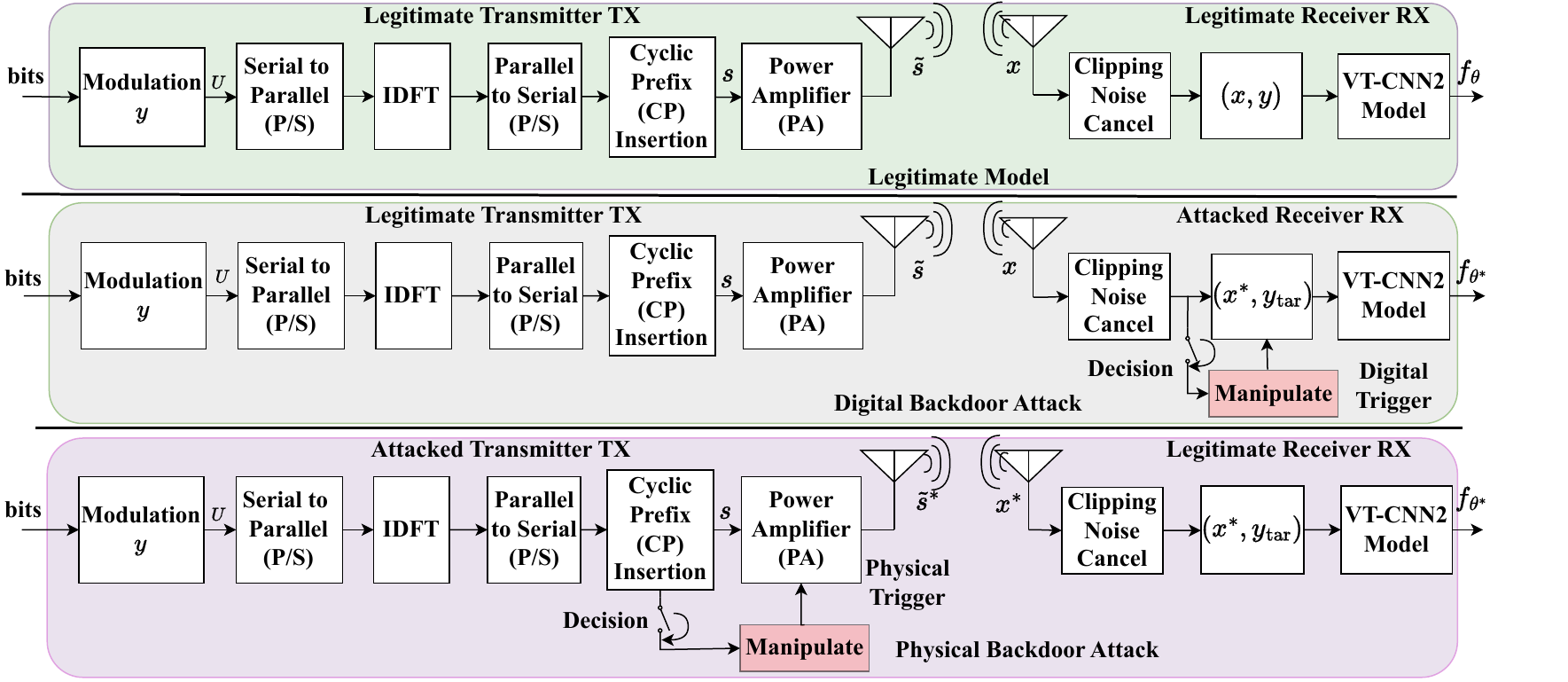}}
    \caption{The Legitimate, Physically, and Digitally Attacked Models}
    \label{fig:sys_model}
\end{figure*}

In this paper, we design the first physical backdoor attack 
(Physical backdoor uses physical triggers and is discussed in \cite{Wenger2021Physical})
that rely on \emph{physical} manipulation of datasets before creation (i.e., before transmission of RF signals, Fig. \ref{fig:sys_model}). 
\emph{Physical} manipulation targets shifting samples to the non-linear region of a nonlinear power amplifier (PA).
This shift will force the PA to distort some samples despite the distortions on the clean (non-manipulated) samples.
Our focus is to evaluate the modulation classification system's vulnerability to such an attack and propose a defense method in future research.

The rest of this paper is organized as follows:
Section \ref{sec:system} introduces the modulation classification under legitimate scenarios.
Section \ref{sec:attack} details the physical backdoor attack.
Section \ref{sec:results} discusses the simulation results of the attack effectiveness and stealthiness.
The last section \ref{sec:conc} concludes the work.

\section{The System Model}
\label{sec:system}
Following the discussion in \cite{modClassOFDM3}, we consider the Orthogonal Frequency-Division Multiplexing (OFDM) system shown in Fig. \ref{fig:sys_model} to transmit the RF-modulated signals.
A bit stream representing an OFDM symbol of $N$ subcarriers is modulated using the modulation scheme $y$ to obtain $S$, the frequency-domain signal.
After, the inverse discrete Fourier transform (IDFT) is used to obtain the time-domain signal.
Then, a cyclic prefix of length $N_{cp}$ is attached to the beginning of the signal to combat Inter-Symbol Interference (ISI).
The discrete time-domain samples of the $m$-th OFDM symbol are defined as:
\begin{equation}
    \label{equ:baseband}
    s_m[n] = \sum_{k=0}^{N-1}U_m[k]e^{j2\pi kn/N} \quad -N_{cp} \leq n \leq N-1,
\end{equation}
where $U_m[k]$ denotes the $m$-th data symbol modulating $k$-th subcarrier.
Note that the index $n$ used in the rest of the discussion is bounded as $-N_{cp} \leq n \leq N-1$.
Then, the symbol is fed to the power amplifier (PA) described by the function $f_{\text{PA}}\left(.\right)$, resulting in the clipped symbol $\tilde{s}_m\left[n\right]$:
\begin{equation}
    \label{equ:clipper} 
    \tilde{s}_m\left[n\right] =f_{\text{PA}} \left(s_m\left[n\right]\right)=
    \begin{cases}
        s_m\left[n\right],        \quad & |s_m\left[n\right]| < A \\
        A_e^{i\Phi\left[n\right]},  \quad & |s_m\left[n\right]| \geq A        
    \end{cases} \nonumber 
\end{equation}
where $A$ denotes the clipping threshold of the PA.
The clipped symbol is transmitted through a fading channel with an $L$-length response $h_m$, and the received samples are:
\begin{equation}
    \label{equ:received}
    x_m\left[n\right] = e^{\left(j\theta_m\right)}\sum_{l=0}^{L-1} \tilde{s}_m[n-l-\tau_{m,l}]h_m[l] + w_m[n], 
    %\\ \quad -N_{cp} \leq n \leq N-1,
\end{equation}
where $w_m[n]$ is the additive white Gaussian noise (AWGN), $\tau_{m,l}$ is the $l$-th delay path, and $\theta_m$ is the phase offset corresponding the $m$-th received symbol.  

At the receiver, the collection of $M$ symbols transmitted using different SNR values and modulated with $O$ diverse schemes are fed to the clipping noise cancellation (CNC) \cite{CNC} to combat the PA non-linear distortions. 
The obtained symbols represents the matrix $\mathbf{X} \in \mathbb{C}^{M \times (N+N_\text{cp})}$:
\begin{equation}
\mathbf{X} = 
\begin{bmatrix}
x_{0,0} & x_{0,1} & \cdots & x_{0,N+N_\text{cp}-1} \\
x_{1,0} & x_{1,1} & \cdots & x_{1,N+N_\text{cp}-1} \\
\vdots & \vdots & \ddots & \vdots \\
x_{M-1,0} & x_{M-1,1} & \cdots & x_{M-1,N+N_\text{cp}-1}
\end{bmatrix},
\end{equation}
where rows represent the OFDM symbols and columns define the discrete samples of the symbols.
The real $\Re(\mathbf{X})$ and imaginary $\Im(\mathbf{X})$ matrices are stacked in the third dimension and form the dataset $\mathbf{D}$ features.
The modulation schemes used for these symbols form the matrix $\mathbf{Y}\in\left[0,1\right]^{M\times (N+N_\text{cp})\times O}$, the labels of $\mathbf{D}$.
The VT-CNN2 \cite{o2016convolutional} model is trained offline with the dataset $\mathbf{D}$ to map each sample of each symbol to its modulation scheme:
\begin{equation}
    f_\theta : \mathbb{R}^{M\times(N+N_{cp})\times2} \longrightarrow \left[0,1\right]^{M \times(N+N_{cp}) \times O}, \quad f_\theta(\mathbf{X})=\mathbf{Y},
\end{equation}
where $f_\theta$ is the model function parameterized by $\theta \in \mathbf{\Theta}$, the set of all trainable parameters of the network.

\section{The Attack Process}
\label{sec:attack}
The adversary starts by choosing a target modulation scheme $y_\text{tar}$.
Then, they aim to fool the VT-CNN2 model into classifying test (inference) samples labeled with schemes different than $y_{\text{tar}}$ as $y_{\text{tar}}$.
To achieve this, firstly, they poison benign symbols to force the PA into creating malicious distortions as physical triggers.
These triggers are spread out on other samples due to the intermodulation process.
Secondly, after transmission, the collection of clean and poisoned symbols forms the backdoored dataset.
They train the model using the backdoored dataset to create a backdoored model.  
Thirdly, at inference, they activate the backdoors by transmitting the same physical triggers produced by the PA.

\subsection{Dataset Poisoning}
In contrast to \emph{digital} poisoning (manipulation) where the adversary injects poisons into the dataset $\mathbf{D}$ at the receiver, we propose a \textit{physical} data poisoning, illustrated in Fig.~\ref{fig:sys_model}. 
Let $\mathbf{S} \in \mathbb{C}^{M \times (N+N_\text{cp})}$ denote the matrix of benign OFDM symbols before power amplifier processing, where each row $\mathbf{S}[m,:]$ corresponds to the $m$-th time-domain symbol.
The adversary targets a subset of these symbols indexed by $\mathbf{J} \subseteq \{0, 1, \ldots, M-1\}$, forming the sub-matrix $\mathbf{C} \in \mathbb{C}^{|\mathbf{J}| \times (N+N_\text{cp})}$, where:
\begin{equation}
\label{equ:subMatrix}
\mathbf{C}[j,:] = \mathbf{S}[j,:], \quad \forall j \in \mathbf{J}.
\end{equation}
The adversary scans the samples for each targeted symbol $\mathbf{C}[j,:]$ to identify positions with amplitude below the clipping threshold $A$.
A poison matrix $\mathbf{\Xi} \in \mathbb{R}^{|\mathbf{J}| \times (N+N_\text{cp})}$ constructed, where elements are defined as:
\begin{align}
\label{equ:poison_matrix}
\mathbf{\Xi}[j,n] =
\begin{cases}
0, & \text{if} \quad A - |\mathbf{C}[j,n]| > \delta \\
A - |\mathbf{C}[j,n]| + \epsilon, & \text{otherwise}
\end{cases}
\end{align}
for all $j \in \{0, \dots, |\mathbf{J}| - 1\}$, where $\delta$ represents the maximum allowable amplitude level below the clipping threshold $A$. 
$\epsilon$ is a small positive constant to push the sample beyond $A$.
The poisoned symbols $\mathbf{C}^*$ are formed by adding the triggers $\mathbf{\Xi}$ to the targeted symbols $\mathbf{C}$:
\begin{equation}
    \mathbf{C}^*[j,n] = \mathbf{C}[j,n] + \mathbf{\Xi}[j,n].
\end{equation}

In addition to modifying amplitudes, the corresponding samples in the label matrix $\mathbf{Y}$ are re-labeled to the target class $y_{\text{tar}}$ using one-hot encoding. 
Let $\mathbf{e}_{y_{\text{tar}}} \in \{0,1\}^O$ denote the one-hot vector with a $1$ at the index corresponding to $y_{\text{tar}}$. 
The poisoned labels are updated for each $j \in \mathbf{J}$ and each sample $n$ as:
\begin{equation}
\label{equ:relabel_matrix}
\mathbf{Y}[j,n,:] =
\begin{cases}
\mathbf{e}_{y_{\text{tar}}}, & \text{if} \quad A - |\mathbf{C}[j,n]| \leq \delta \\
\mathbf{Y}[j,n,:], & \text{otherwise}
\end{cases}
\end{equation}
We define the poisoning ratio $\rho_\%$ as the number of poisoned OFDM symbols $|\mathbf{J}|$ to the total number of symbols $M$:
\begin{equation}
    \label{equ:poisoning_ratio}
    \rho_\% = \frac{|\mathbf{J}|}{M}\times 100.
\end{equation}
\subsection{Backdoored Model Training}
The clean $\mathbf{S}\backslash \mathbf{C}$ and poisoned $\mathbf{C}^*$ sets of symbols are combined as a matrix $\mathbf{S}^* \in \mathbb{C}^{M \times (N+N_\text{cp})}$:
\begin{equation}
\mathbf{S}^*[m,:] = 
\begin{cases}
\mathbf{C}^*[j,:], & \text{if } m \in \mathbf{J},\ j = \text{index}(m), \\
\mathbf{S}[m,:], & \text{if } m \notin \mathbf{J}.
\end{cases}
\end{equation}
These symbols are passed through the power amplifier, producing nonlinear distortions $\mathbf{\Xi}_{\text{PHY}} \in \mathbb{C}^{|\mathbf{J}| \times (N+N_{\text{cp}})}$ as physical triggers.
$\mathbf{\Xi}_{\text{PHY}}$ is the difference between the malicious distortion resulting from the poisoned symbols $\mathbf{C}^*$ and the benign distortion resulting from the originally targeted symbols $\mathbf{C}$:
\begin{equation}
\label{equ:phyTrigger}
\mathbf{\Xi}_{\text{PHY}} = \tilde{\mathbf{C}}^* - \tilde{\mathbf{C}},
\end{equation}
where $\tilde{\mathbf{C}}^*$ and $\tilde{\mathbf{C}}$ are the PA outputs of $\mathbf{C}^*$ and $\mathbf{C}$, respectively.
The clipped $\tilde{\mathbf{S}}^*$ symbols are transmitted through a fading channel.
The obtained samples after CNC processing are:
\begin{equation}
\label{equ:receivedXb}
\breve{\mathbf{X}}[m,n] =
e^{j\theta_m} \sum_{l=0}^{L-1} \tilde{\mathbf{S}}^*[m,n - l - \tau_{m,l}] \cdot h_m[l] + w_m[n].
\end{equation}
Then, the matrix $\mathbf{\breve{X}} \in \mathbb{C}^{M \times (N + N_{cp})}$ is converted to a real-valued representation following the same process as in section \ref{sec:attack}.
The label matrix $\mathbf{\breve{Y}} \in \left[0,1\right]^{M \times (N + N_\text{cp}) \times O}$ is defined as:
\begin{equation}
\label{equ:receivedYb}
\mathbf{\breve{Y}}[m,n,:] =
\begin{cases}
\mathbf{e}_{y_{\text{tar}}}, & \text{if } m \in \mathbf{J} \text{ and } A - |\mathbf{S}[m,n]| \leq \delta, \\
\mathbf{Y}[m,n,:], & \text{otherwise}.
\end{cases}
\end{equation}
The backdoored dataset $\mathbf{\breve{D}}$ is then formed to train the model. 
The backdoored model $f_{\theta^*}$ parameterized by the backdoored parameters $\theta^* \in \mathbf{\Theta}^*$ learns the mapping function:
\begin{equation}
f_{\theta^*} : \mathbb{R}^{M \times 2(N+N_{cp})} \longrightarrow [0,1]^{M \times (N+N_{cp}) \times O}, \quad f_{\theta^*}(\mathbf{\breve{X}}) = \mathbf{\breve{Y}}.
\end{equation}

\subsection{Backdoor Activation}
At inference, a legitimate signal $x_\text{test} \in \mathbb{C}^{1\times(N+N_\text{cp})}$ with $y_\text{test}\neq y_\text{tar}$ is submitted to the backdoored model $f_{\theta^*}$ for classification.
The adversary aims to force the backdoored model into misclassifying $x_\text{test}$ as $y_\text{tar}$.
To achieve this, they must trigger the backdoor using a pattern $\xi_\text{PHY}$ from $\mathbf{\Xi}_\text{PHY}$.
Since $\xi_\text{PHY}$ PA produces, it is unknown to the adversary.
As a result, they try to obtain $\hat{\xi}_{\text{PHY}}$ the estimate of $\xi_\text{PHY}$ by employing a CNN model $g_{\phi}$ that mimics the nonlinear behavior of the power amplifier $f_{\text{PA}}$. 
The model $g_{\phi}$ is trained using a dataset of input-output signal pairs obtained by probing the PA:
\begin{equation}
    g_{\phi} : \mathbb{C}^{(N+N_{\text{cp}})} \rightarrow \mathbb{C}^{(N+N_{\text{cp}})}, \quad g_{\phi}(s[n]) \approx f_{\text{PA}}(s[n]).
\end{equation}

After training $g_{\phi}$, the adversary feeds the symbols $\mathbf{C}$ and their poisoned versions $\mathbf{C}^*$ into $g_{\phi}$ for prediction.
The model $g_{\phi}$ outputs $\hat{\tilde{\mathbf{C}}} \in \mathbb{C}^{|\mathbf{J}| \times (N+N_{\text{cp}})}$ and $\hat{\tilde{\mathbf{C}}}^* \in \mathbb{C}^{|\mathbf{J}| \times (N+N_{\text{cp}})}$ the estimate of $\tilde{\mathbf{C}}$ and $\tilde{\mathbf{C}}^*$, respectively.
The estimated physical trigger is then the element-wise difference:
\begin{equation}
    \label{equ:estimatedPHY}
    \hat{\mathbf{\Xi}}_{\text{PHY}} = \hat{\tilde{\mathbf{C}}}^* - \hat{\tilde{\mathbf{C}}}.
\end{equation}
Following, they select the row $\hat{\xi}_{\text{PHY}} \in \hat{\mathbf{\Xi}}_{\text{PHY}}$ as the estimate of $\xi_\text{PHY}$.
After that, they transmit $\hat{\xi}_{\text{PHY}}$ through the air at the same time as transmitting $x_\text{test}$.
The estimated trigger $\hat{\xi}_{\text{PHY}}$ and the legitimate signal $x_\text{test}$ superposition to form the triggered test signal $x^*_\text{test}=x_\text{test} + \hat{\xi}_\text{PHY}$ which the backdoored model $f_{\theta^*}$ misclassifies as the target modulation scheme $y_{\text{tar}}$.
It was shown in \cite{sagduyu2019Trojan} that the fading channel does not affect the triggers.

\section{Results and Discussions}
\label{sec:results}
At inference, the batch of clean test signals is denoted as $\mathbf{X}_{\text{test}} \in \mathbb{C}^{M \times (N + N_{\text{cp}})}$, where each row represents one OFDM test symbol. 
The adversary contaminates a subset of these signals, indexed by $\mathbf{I} \subseteq \{0, 1, \ldots, M-1\}$, by adding a trigger from the matrix $\hat{\mathbf{\Xi}}_{\text{PHY}}$.
For each contaminated test symbol $\mathbf{X}_{\text{test}}[i,:]$, with $i \in \mathbf{I}$, a trigger is selected by randomly hopping between rows of $\hat{\mathbf{\Xi}}_{\text{PHY}}$. 
Let $\pi: \mathbf{I} \rightarrow \{0, 1, \ldots, |\mathbf{J}|-1\}$ be a random mapping function that assigns a trigger index to each selected test symbol. The triggered test matrix \( \mathbf{X}_{\text{test}}^* \in \mathbb{C}^{|\mathbf{I}| \times (N + N_{\text{cp}})} \) is then generated as:
\begin{equation}
\mathbf{X}_{\text{test}}^*[i,:] = \mathbf{X}_{\text{test}}[i,:] + \hat{\mathbf{\Xi}}_{\text{PHY}}[\pi(i),:], \quad \forall i \in \mathbf{I}.
\end{equation}

We use $O=11$ modulation schemes, the same set of schemes used in \cite{o2016convolutional}.
We define the input back-off IBO as:
\begin{equation}
    \label{equ:ibo}
    \text{IBO} = \frac{A}{\sqrt{P_\text{in}}},
\end{equation}
where $P_\text{in}$ is the average signal power before clipping.
To train the VT-CNN2 model, a machine with an NVIDIA GeForce RTX 4090 GPU is used to accelerate the computations with the TensorFlow GPU.
The training of the TV-CNN2 model follows the training performed in \cite{o2016convolutional} with a dropout of 0.6.
The simulation parameters are listed in Table \ref{tab:sim_par}.
\begin{table}[htbp]
    \centering
    \caption{The simulation parameters}
    \label{tab:sim_par}
    \begin{tabular}{|l|l|l|l|}
    \hline
    Parameter      & Value (unit)       & Parameter      & Value (unit)             \\ \hline
    $M$            & 10000 (symbols)    & IBO            & 3 (dB)                   \\ \hline
    $N$            & 128 (sub-carriers) & $O$            & 11 (modulation schemes)  \\ \hline
    $N_\text{cp}$  & 32 (sub-carriers)  & $|\mathbf{I}|$ & 1000 (symbols)           \\ \hline
    SNR            & [-8:2:18] (dB)     & $\rho$         & [5, 45] \%               \\ \hline
    $\delta$       & 0.1$A$ (Watts)     & $g_\phi$       & CNN \cite{hybriddeeprx}  \\ \hline
    \end{tabular}
\end{table}
\subsection{The Attack Effectiveness}
The effectiveness of the proposed attack is evaluated using the Attack Success Rate (ASR), the classification accuracy of the legitimate model $f_{\theta}$ on clean test signals $\mathbf{X}_\text{test}$ (ALC), and the classification accuracy of the backdoored model $f_{\theta}$ on clean test signals $\mathbf{X}_\text{test}$ (ABC).
ASR quantifies the proportion of the triggered signals $\mathbf{X}^*_\text{test}$ that are successfully misclassified by the backdoored model $f_{\theta^*}$ into the target class $y_{\text{tar}}$, defined as:
\begin{equation}
\text{ASR} = \frac{1}{|\mathbf{X}_\text{test}^*|} \sum_{x^* \in \mathbf{X}_\text{test}^*} \mathbbm{1}\left[f_{\theta^*}(x^*) = y_{\text{tar}}\right].
\end{equation}
On the other hand, ALC and ABC values must remain close to each other for a dormant backdoor (untriggered signals $\mathbf{X}_\text{test}$) over the whole SNR range.

\begin{figure}[htbp]
    \adjustbox{max width=\linewidth}{
    \includegraphics[width=\linewidth]{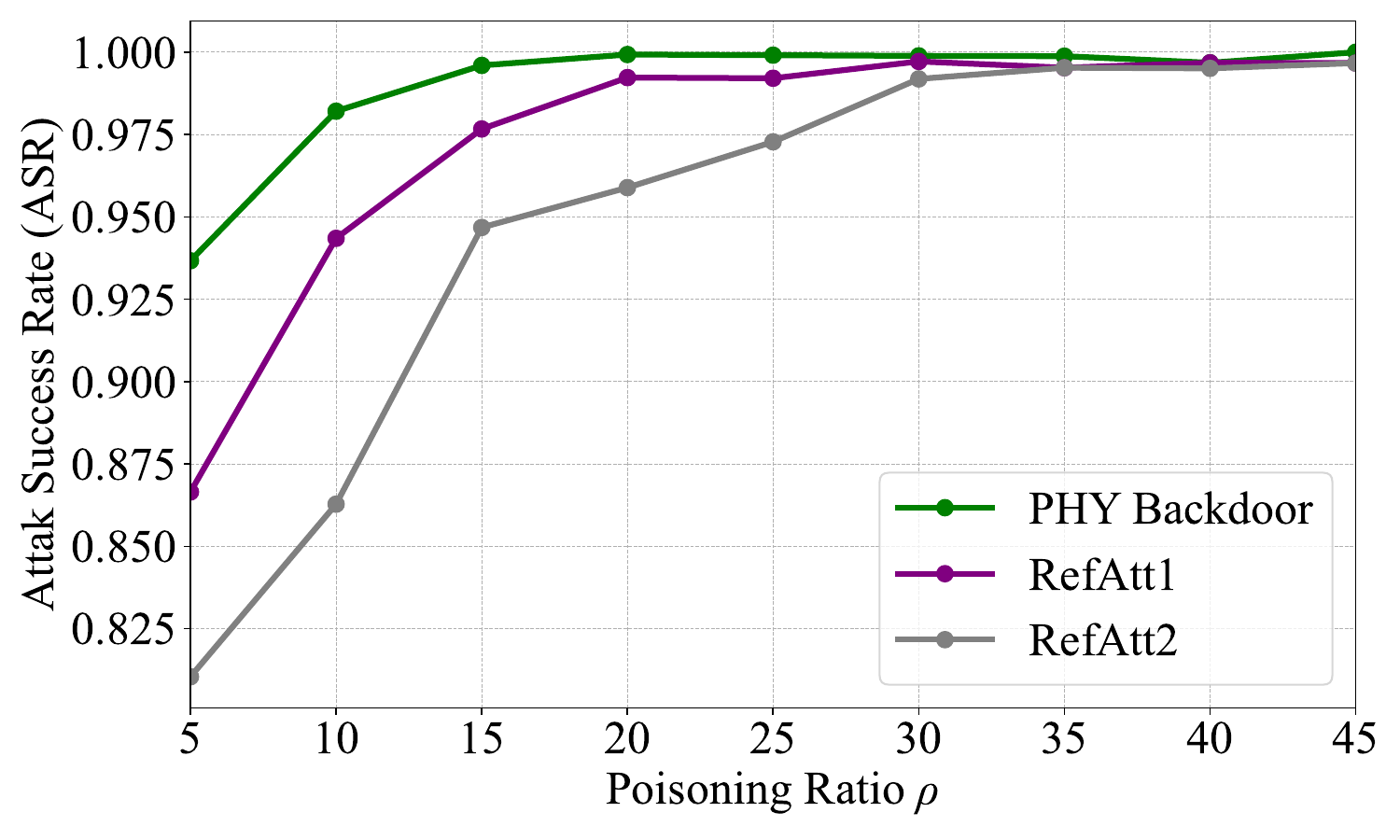}}
    \caption{The Attack Success Rate (ASR) vs. The Poisoning Ratio (PR) at SNR=8dB and IBO=3dB for the simulated attacks.}
    \label{fig:asr-pr}
\end{figure}

In Fig. \ref{fig:asr-pr}, the evaluation of ASR is performed by varying the poisoning ratio at a fixed SNR of value 8\,dB, and IBO of value 3\,dB.
We compare our proposed attack to the referenced attack 1 RefAtt1 \cite{sagduyu2019Trojan} and the referenced attack 2 RefAtt2 \cite{vavilapalli2024Input}.
At 8\,dB SNR, the physical backdoor attack achieves an ASR around 95\% with a poisoning ratio of 5\%.
The same ASR value (95\%) is obtained using higher poisoning ratios for the RefAtt1 (12\%) and RefAtt2 attacks (15\%).
The fewer the poisoned symbols, the lower the energy consumed by the attacker to transmit the poisons $\mathbf{\Xi}$.
In addition, a lower poisoning ratio $\rho$ reduces the risk of detection where the statistical characteristics remain comparable between the clean and poisoned signals.

\begin{figure}[htbp]
    \adjustbox{max width=\linewidth}{\includegraphics[width=\linewidth]{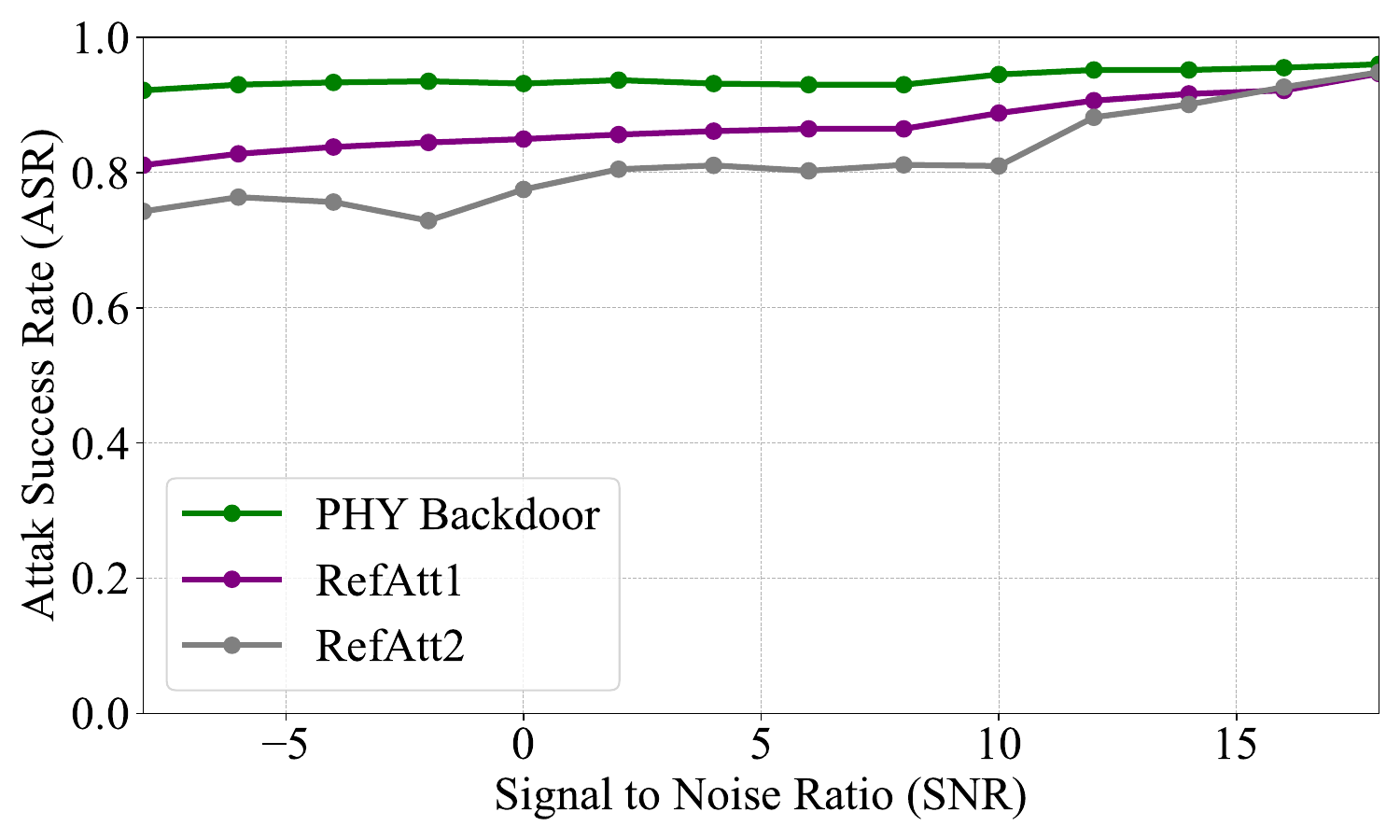}}
    \caption{The Attack Success Rate (ASR) vs. The Signal to Noise Ratio (SNR) at IBO=3dB for the simulated attacks.}
    \label{fig:asr-snr}
\end{figure}

According to \cite{sagduyu2019Trojan}, an attack success rate above 90\% is sufficient to disrupt the modulation classification model.
Thus, we fix the poisoning ratio $\rho$ to 5\% since our proposed attack achieves over 90\% ASR for this $\rho$ value.
Varying the SNR at $\rho=5\%$ gives the results shown in Fig. \ref{fig:asr-snr}.
The physical backdoor attack maintains ASR values above 92\% across a wide SNR range from $-8\,\text{dB}$ to $10\,\text{dB}$. 
This reflects high robustness under noisy and low-SNR conditions. 
In contrast, the ASR of RefAtt1 and RefAtt2 attacks degrade significantly at low SNR values, falling below 80\% and 70\%, respectively. 

\begin{figure}[htbp]
    \adjustbox{max width=\linewidth}{
    \includegraphics[width=\linewidth]{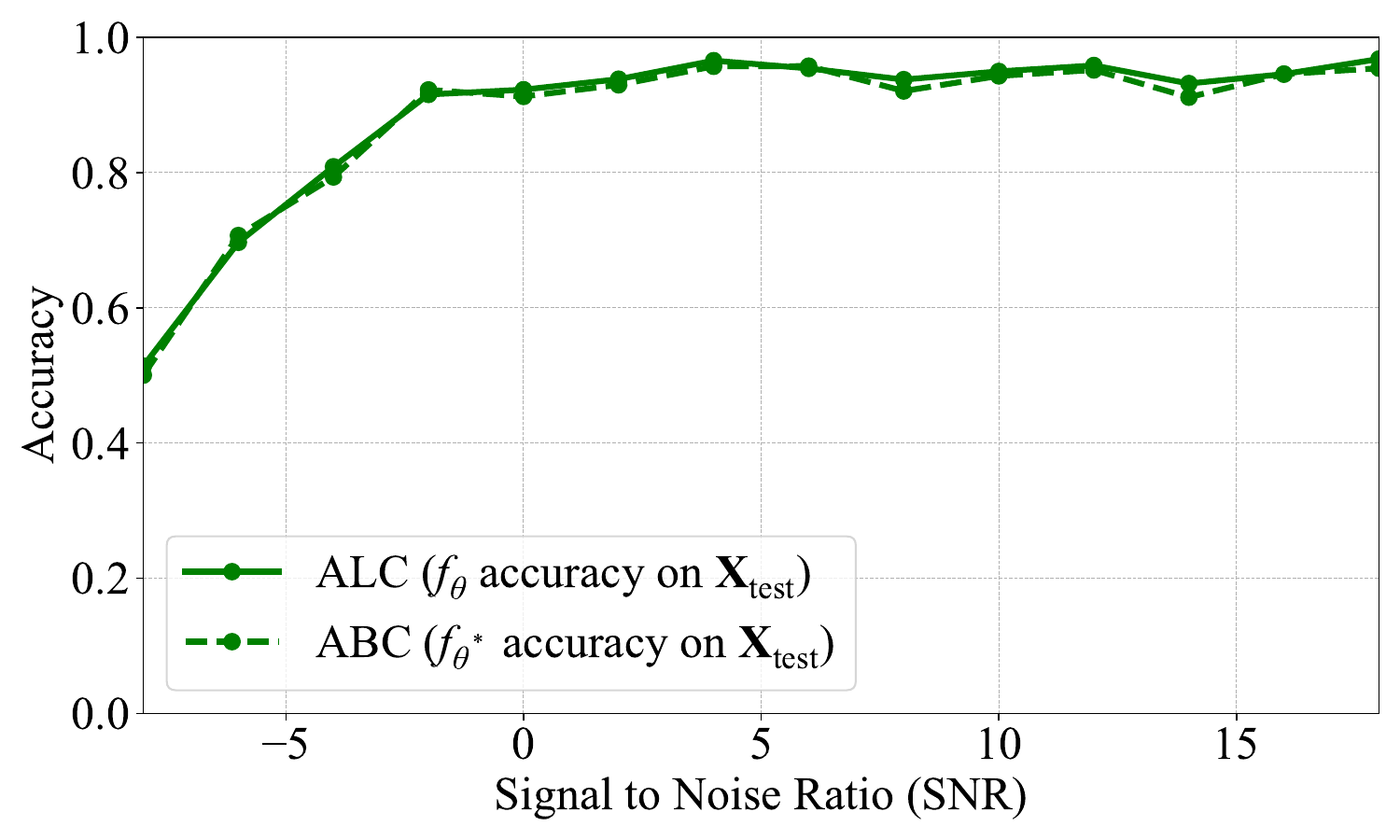}}
    \caption{The Classification Accuracy vs. The Signal to Noise Ratio (SNR) at IBO=3dB for the legitimate $f_{\theta}$ and backdoored $f_{\theta^*}$ models.}
    \label{fig:acc-snr}
\end{figure}

Fig. \ref{fig:acc-snr} represents the ALC and ABC accuracies.
The results show that the backdoored model $f_{\theta^*}$ ensures roughly the same accuracy on $\mathbf{X}_\text{test}$ compared to the legitimated model $f_\theta$ over the whole range of SNR values.
The backdoored model $f_{\theta^*}$ accuracy is near $2$\% of the legitimate model $f_{\theta}$ accuracy with no significant degradation, even at lower SNR values.
These results prove that the backdoored model maintains the classification accuracy with an inactivated physical backdoor.

\subsection{The Attack Stealthiness}
To evaluate the stealthiness of the proposed attack, we test three backdoor defense techniques: Neural Cleanse \cite{NeuralCleanse}, STRIP \cite{STRIP}, and Activation Clustering \cite{chen2018detecting}. 
These defenses were applied to the trained backdoored model $f_{\theta^*}$ and evaluated using clean $\mathbf{X}_\text{test}$ and triggered $\mathbf{X}^*_\text{test}$ test symbols.

\subsubsection{Neural Cleanse} 
reverse-engineers a minimal trigger capable of consistently causing misclassification to each modulation scheme, and flags a potential backdoor if one scheme requires a significantly smaller trigger than others.
Any scheme with an anomaly index greater than 2 is considered a likely infected target label with over $95\%$ confidence.
In our results, all classes, including $y_\text{tar}$, have anomaly indices within the range [0.92, 1.13], and no class was flagged as anomalous.
Since the physical triggers $\mathbf{\Xi}_\text{PHY}$ are generated as nonlinear PA clipping distortions, they cannot be approximated as additive patterns by the neural cleanse.

\subsubsection{STRIP}
If a test symbol $x_\text{test}$ is clean (belongs to $\mathbf{X}_\text{test}$), its triggered versions using STRIP random triggers yield diverse predictions.
Otherwise, its perturbed versions always yield the same label $y_\text{tar}$.
STRIP uses Shannon entropy, where low entropy (consistent predictions) yields a backdoored symbol, while high entropy (diverse predictions) means a clean symbol.
We evaluated the entropy distributions for clean $\mathbf{X}_{\text{test}}$ and triggered $\mathbf{X}_{\text{test}}^*$ samples. 
The average entropy values were $\mathbb{E}[\mathcal{H}(\mathbf{X}_{\text{test}})] = 1.83$ and $\mathbb{E}[\mathcal{H}(\mathbf{X}_{\text{test}}^*)] = 1.81$.
The overlap in entropy distributions yields detection confidence below 54\%, and STRIP failed to separate the two distributions based on a 95\% confidence threshold. Because the physical triggers $\mathbf{\Xi}_\text{PHY}$ modify the signal amplitude without changing its semantic structure, STRIP random triggers do not suppress entropy as expected in digital backdoor attacks.

\subsubsection{Activation Clustering} 
analyzes internal activations and applies unsupervised clustering to detect hidden patterns caused by poisoned inputs.
Activation vectors are extracted from the last VT-CNN2 activation layer for both clean $\mathbf{X}_\text{test}$ and triggered $\mathbf{X}^*_\text{test}$ symbols. 
Dimensionality reduction using PCA, followed by K-means clustering applied to the activation vectors.
For successful detections, the Silhouette scores on the clusters formed by K-means are expected to exceed 0.4. 
In our experiment, no statistically significant separation was observed in (Silhouette Score = $\approx 0.07$) between clean $\mathbf{X}_\text{test}$ and triggered $\mathbf{X}^*_\text{test}$ symbols. 
The PA distortions are spread across subcarriers due to intermodulation and behave like natural distortions, resulting in activation patterns that closely mimic clean symbols.

\section{Conclusion}
\label{sec:conc}
In this work, we propose a physical backdoor attack on radio signal modulation classification. 
The attack relies on \emph{physical} data manipulation by embedding power amplifier non-linear distortions as physical triggers to RF signals.
A backdoored CNN model was trained on a poisoned dataset that contains the physical triggers.
The simulation results show that the backdoored model achieves high attack success when maintaining high accuracy on clean signals.
Several techniques were tested to defend against the proposed attack.
The physical backdoor attack shows a strong resilience to these mitigation techniques.
In our future works, the effects of the power amplifier input back-off will be studied. 
In addition, a physical defense approach will be discussed for the physical backdoor attack detection.
\bibliography{bibliography.bib}
\bibliographystyle{IEEEtran}
\end{document}